# Effective Hamiltonian for a half-filled tetramerized ionic-Hubbard chain

M. Menteshashvili, N. Chachava and G.I. Japaridze

*Andronikashvili Institute of Physics, 6 Tamarashvili St, 0177 Tbilisi, Georgia*

**Abstract.** *We derive an effective spin Hamiltonian for the one-dimensional half-filled tetramerized ionic-Hubbard model in the limit of strong on-site repulsion. We show that the effective Hamiltonian which describes the low-energy spin sector of the model is a spin S=1/2 Heisenberg Hamiltonian with alternating nearest-neighbour exchange.*

**Introduction.** The Hubbard model is the most commonly used model for studying correlated fermions on a lattice [1]. The usual and simplest version of the Hubbard model, given by the Hamiltonian

$$\hat{H} = -t\sum_{i,j,\sigma} N_{ij}\hat{c}^+_{i\sigma}\hat{c}_{j\sigma} + U\sum_{i=1}^{N}\hat{n}_{i\uparrow}\hat{n}_{i\downarrow}, \qquad (1)$$

is characterized by two energy scales defined by $t$ and $U$, where $t$ is the electron hopping amplitude between nearest-neighbour sites and $U > 0$ is the on-site Coulomb repulsion energy. In Eq. (1), $\hat{c}^+_{i\sigma}$ ($\hat{c}_{i\sigma}$) is the creation (destruction) operator of an electron with spin $\sigma$ on site $i$, $\hat{n}_{i\sigma} = \hat{c}^+_{i\sigma}\hat{c}_{i\sigma}$ is the number operator and $N_{ij} = 1$ if $i, j$ are nearest-neighbour sites and is zero otherwise.

One of the virtues of the Hubbard Hamiltonian is that at half-filling and strong repulsion ($U \gg t$), it provides a very simple picture of the Mott insulator (MI) [2]. In the MI state, the charge excitations are suppressed by the large (Mott) gap and the low energy sector of excitations is fully represented by the spin degrees of freedom. There exist several very elegant methods for deriving the effective low-energy spin Hamiltonian for the Hubbard model [3-6]. Among the most widely used is the so-called canonical transformation (CT) method [3,4], which introduces a unitary transformation that "rotates" $\hat{H}$ into an effective spin-only Hamiltonian $\hat{H}_{eff}$ and the corresponding state vectors into the restricted spin-only subspace. In performing a CT, the high-energy (charge) degrees of freedom are taken into account in the effective low-energy theory through non-local effective interactions. The true ground state eigenvector of the theory is in essence "rotated" to coincide with the ground state of the subspace of the effective low-energy theory. In the case of standard Hubbard model (1), the low-energy sector is described by the effective spin $S = 1/2$ Heisenberg model with nearest-neighbour spin exchange $J = t^2/U$ [7].

Recently, the extended version of the half-filled Hubbard chain with alternating on-site energies $\pm \Delta_0/2$, known as the ionic Hubbard model (IHM), has attracted much of the current interest [8-17]. The model describes an atomic chain built as an alternating sequence of different ("A" and "B") atoms, and the additional energy scale $\Delta_0$ marks the difference between single-electron energies on neighbouring sites. Although initially the IHM has been proposed to describe the neutral-ionic transition in charge-transfer organic crystals [8] and ferroelectric perovskites [9], the increased current interest in this model is mostly motivated by its rich ground state phase diagram, which exhibits, with the increasing Coulomb repulsion, a transition from a band insulating phase at $U \ll \Delta_0$ into the MI phase at $U \gg \Delta_0$ [10-17]. In complete agreement with the known properties of the MI state, the low-energy spin sector of the half-filled IHM at $U \gg \Delta_0, t$ is gapless and is described by the Hamiltonian of a spin $S = 1/2$ Heisenberg chain with nearest-neighbour spin exchange $J = t^2/U(1 - \Delta_0^2/U^2)$ [8].

In this paper we derive the effective spin Hamiltonian for the half-filled IHM with tetramerized ionic potential corresponding to the AABB-type chain. We use the CT method and demonstrate that, in marked contrast with previously studied cases, the low-energy spin sector of the tetramerized IHM chain at $U \gg \Delta_0, t$ is *gapped* and is described by the Hamiltonian of a spin $S = 1/2$ Heisenberg model with alternating spin exchange.

The Hamiltonian we consider is given by the expression

$$\hat{H} = -t \sum_{i,j,\sigma} N_{ij} \hat{c}^+_{i\sigma} \hat{c}_{j\sigma} + U \sum_{i=1}^{N} \hat{n}_{i\uparrow} \hat{n}_{i\downarrow} + \frac{\Delta_0}{\sqrt{2}} \sum_{i,\sigma} \cos\left(\frac{\pi}{4} + \frac{\pi}{2}i\right) \hat{n}_{i\sigma} \equiv \hat{T} + \hat{V} + \hat{\Delta} \qquad (2)$$

and contains alongside with the standard hopping and on-site repulsion terms the ionic term which describes the modulation of the lattice potential. We restrict our consideration by the strong coupling limit assuming $U \gg t, \Delta_0$ and adopt the CT method developed in Ref. [4] to derive the effective low-energy spin Hamiltonian.

**The CT Method**. Contrary to the case of the ordinary Hubbard model (1) where at half-filling the sub-bands can be classified solely by the total number of doubly occupied sites (doublons) $N_d$ [4], in the case of IHM we have to deal with a system where each sub-band is characterized by two different numbers: by the overall number of doublons $N_d$ and the difference between the numbers of electrons on low on-site energy ("A") sites and high on-site energy ("B") sites $N_e^B - N_e^A$. The hopping term in some cases does mix states from different sub-bands and in some cases it does not. The mixing of the sub-bands can be avoided by introducing suitable linear combinations of the uncorrelated basic states (at $U > 0$ and half-filling it is sufficient to achieve the "non-mixing" of only those sub-bands which differ by the number of doublons). The $\hat{S}$ matrix for this transformation, and the transformed Hamiltonian $\hat{H}' = e^{i\hat{S}} \hat{H} e^{-i\hat{S}}$, are generated by an iterative procedure. As introduced in Refs. [3,4], the CT relies on the separation of the kinetic part $\hat{T}$ into three terms: $\hat{T}_1$, which increases the number of doubly occupied sites by one, $\hat{T}_{-1}$, which decreases the number of doubly occupied sites by one, and $\hat{T}_0$, which leaves this number unchanged ($\hat{h}_{i\sigma} \equiv 1 - \hat{n}_{i\sigma}$, $\bar{\sigma} \equiv -\sigma$):

$$\hat{T}_1 = -t \sum_{i,j,\sigma} N_{ij} \hat{n}_{i\bar{\sigma}} \hat{c}^+_{i\sigma} \hat{c}_{j\sigma} \hat{h}_{j\bar{\sigma}} ; \tag{3a}$$

$$\hat{T}_{-1} = -t \sum_{i,j,\sigma} N_{ij} \hat{h}_{i\bar{\sigma}} \hat{c}^+_{i\sigma} \hat{c}_{j\sigma} \hat{n}_{j\bar{\sigma}} ; \tag{3b}$$

$$\hat{T}_0 = -t \sum_{i,j,\sigma} N_{ij} (\hat{n}_{i\bar{\sigma}} \hat{c}^+_{i\sigma} \hat{c}_{j\sigma} \hat{n}_{j\bar{\sigma}} + \hat{h}_{i\bar{\sigma}} \hat{c}^+_{i\sigma} \hat{c}_{j\sigma} \hat{h}_{j\bar{\sigma}}) . \tag{3c}$$

In addition, in the case of tetramerized IHM each of these terms is split into several further terms which specify between which kind of sites ("A" or "B") the hopping process occurs:

$$\hat{T}_m = \hat{T}_m^{A \leftarrow A} + \hat{T}_m^{A \leftarrow B} + \hat{T}_m^{B \leftarrow A} + \hat{T}_m^{B \leftarrow B} , \quad m = 0, \pm 1 . \tag{4}$$

Obviously, the hopping processes $\hat{T}_m^{A \leftarrow A}$ and $\hat{T}_m^{B \leftarrow B}$ do not change the ionic energy, while the $\hat{T}_m^{B \leftarrow A}$ ($\hat{T}_m^{A \leftarrow B}$) term increases (reduces) the ionic energy by $\Delta_0$. Below we use the notation $\hat{T}_m^{A \leftarrow A} + \hat{T}_m^{B \leftarrow B} \equiv \hat{T}_m^0$, $\hat{T}_m^{B \leftarrow A} \equiv \hat{T}_m^1$, $\hat{T}_m^{A \leftarrow B} \equiv \hat{T}_m^{-1}$, where the upper index corresponds to the change of the ionic energy. One can easily check that $(\hat{T}_m^\alpha)^+ = \hat{T}_{-m}^{-\alpha}$ and

$$\left[\hat{V} + \hat{\Delta}, \hat{T}_m^\alpha \right] = (mU + \alpha \Delta_0) \hat{T}_m^\alpha \qquad (\alpha = 0, \pm 1) . \tag{5}$$

Using the commutation relations (5), one also obtains that similar relations hold for the products of $k$ ($k \in \mathbb{N}$) $\hat{T}_m^\alpha$ operators $\hat{T}_{m_1}^{\alpha_1}\hat{T}_{m_2}^{\alpha_2}.....\hat{T}_{m_k}^{\alpha_k} \equiv \hat{T}^{[k]}[m,\alpha]$:

$$[\hat{V}+\hat{\Delta},\hat{T}^{[k]}[m,\alpha]] = \left[U\sum_{i=1}^{k} m_i + \Delta_0 \sum_{i=1}^{k} \alpha_i\right]\hat{T}^{[k]}[m,\alpha]. \qquad (6)$$

Let us now start to search for such a unitary transformation $\hat{S}$ which eliminates in the transformed Hamiltonian

$$\hat{H}' = e^{i\hat{S}}\hat{H}e^{-i\hat{S}} = \hat{H} + \frac{1}{1!}[i\hat{S},\hat{H}] + \frac{1}{2!}[i\hat{S},[i\hat{S},\hat{H}]] + \frac{1}{3!}[i\hat{S},[i\hat{S},[i\hat{S},\hat{H}]]] + ... \qquad (7)$$

the undesirable terms corresponding to hops between states with different numbers of doubly occupied sites. To first order, this can be achieved by choosing

$$i\hat{S} = i\hat{S}^{(1)} = \frac{1}{U}(\hat{T}_1^0 - \hat{T}_{-1}^0) + \frac{1}{U+\Delta_0}(\hat{T}_1^1 - \hat{T}_{-1}^{-1}) + \frac{1}{U-\Delta_0}(\hat{T}_1^{-1} - \hat{T}_{-1}^1), \qquad (8)$$

where $i\hat{S}^{(1)}$ has been chosen with the help of commutation relations (5) so that after inserting it into the expansion (4), it will cancel all hopping terms in $\hat{H}$ except $\hat{T}_0^0$ and $\hat{T}_0^{\pm 1}$, which are the terms of the order $t$ which leave $N_d$ unchanged. (In Eq. (8) and onwards, for arbitrary operator $\hat{P}$, $\hat{P}^{(k)}$ denotes an operator which contains terms of the order of up to and including $t^k$, while $\hat{P}^{[k]}$ denotes an operator which is of the order of $t^k$ exclusively.) The indicated choice of $i\hat{S}$ will surely give among the second-order terms ones which mix states from different sub-bands (let us denote them by $\hat{H}'^{[2]}_{und}$), and to eliminate these one needs to add to the transformation matrix $i\hat{S}$ a corresponding term $i\hat{S}^{[2]}$, determined from the condition

$$[i\hat{S}^{[2]},\hat{V}+\hat{\Delta}] = -\hat{H}'^{[2]}_{und}. \qquad (9)$$

Repeating this procedure and making use of the commutation relations (6), after $k$ steps one obtains a transformed Hamiltonian which contains undesirable terms only of the order of $t^{k+1}$:

$$\hat{H}'^{(k+1)} = e^{i\hat{S}^{(k)}}\hat{H}e^{-i\hat{S}^{(k)}}, \qquad (10)$$

where $i\hat{S}^{(k)}$ is determined by the recursive relation $i\hat{S}^{(k)} = i\hat{S}^{(k-1)} + i\hat{S}^{[k]}$, with $i\hat{S}^{[k]}$ chosen from the condition analogous to (9) that it should eliminate all undesirable terms of the order of $t^k$ in the transformed Hamiltonian:

$$[i\hat{S}^{[k]},\hat{V}+\hat{\Delta}] = -\hat{H}'^{[k]}_{und}. \qquad (11)$$

So, up to the order of $t^k$, the transformed Hamiltonian $\hat{H}'$ will contain no terms which mix states from different sub-bands, from where it obviously follows that its ground state and low-energy excitations belong to the sub-band which is characterized by the minimal value of the site-diagonal part $\hat{V} + \hat{\Delta}$ of the original Hamiltonian (2). Recalling that the original IHM Hamiltonian $\hat{H}$ is connected with $\hat{H}'$ via the unitary transformation $\hat{H}' = e^{i\hat{S}} \hat{H} e^{-i\hat{S}}$, the properties of the low-lying levels of $\hat{H}$ and their corresponding states can be directly studied by analyzing those of $\hat{H}'$.

**Effective Hamiltonian**. In the limit of strong Coulomb repulsion $U \gg \Delta_0, t$, the minimum of the "site-diagonal" energy is reached in the subspace $L$ with $N_d = 0$. Since the states in this subspace differ from each other only by the electron spin configurations, operation of the transformed Hamiltonian $\hat{H}'$ in subspace $L$ is equivalent to the operation of a certain spin $S = 1/2$ Hamiltonian. To derive this effective spin Hamiltonian, we first note that, for any state $|\Psi\rangle_L$ from $L$, there are no hops possible without increasing the number of doubly occupied sites, and therefore

$$\hat{T}_{-1}^{\alpha}|\Psi\rangle_L = 0, \quad \hat{T}_0^{\alpha}|\Psi\rangle_L = 0 \quad \text{for} \quad \alpha = 0, \pm 1. \tag{12}$$

It is also evident that $\hat{T}^{[k]}[m, \alpha]|\Psi\rangle_L = 0$, if either of the following conditions is fulfilled:

a) $\quad m_k \neq 1 \text{ or } \sum_{i=n+1}^{k} m_i = 0 \text{ and } m_n \neq 1 \quad (n = 1, 2, ..., k-2);$ \hfill (13a)

b) $\quad \left| \sum_{i=n}^{k} \alpha_i \right| > \sum_{i=n}^{k} m_i, \quad n = 1, 2, ..., k.$ \hfill (13b)

Furthermore, neither of these conditions may hold but the hopping process corresponding to $\hat{T}^{[k]}[m, \alpha]$ operator may still be unrealizable due to the symmetries of the original Hamiltonian or due to the specific structure of the lattice. For example, since the Hamiltonian of the model (2) at half-filling is invariant under the transformation $t \to -t$, $\Delta_0 \to -\Delta_0$ (what can readily be verified using the particle-hole transformation $\hat{c}_{i\sigma} \leftrightarrow \hat{c}_{i\sigma}^+ \Rightarrow \hat{n}_{i\sigma} \leftrightarrow \hat{h}_{i\sigma}$), the transformed Hamiltonian does not contain terms which are proportional to odd powers of $t$, i.e. $\hat{T}^{[2\ell+1]}[m, \alpha]|\Psi\rangle_L = 0$. The relations (12)-(13) allow us to eliminate many terms from the expansion for $\hat{H}'$ in $L$ subspace and to the fourth order of $t$, we arrive at the following expression for the "rotated" Hamiltonian:

$$\hat{H}_L'^{(4)} = -\frac{1}{U}\hat{T}_{-1}^0\hat{T}_1^0 - \frac{1}{U+\Delta_0}\hat{T}_{-1}^{-1}\hat{T}_1^1 - \frac{1}{U-\Delta_0}\hat{T}_{-1}^1\hat{T}_1^{-1} - \frac{\hat{T}_{-1}^0\hat{T}_{-1}^0\hat{T}_1^0\hat{T}_1^0 - 2\hat{T}_{-1}^0\hat{T}_1^0\hat{T}_{-1}^0\hat{T}_1^0}{2U^3} -$$

$$-\frac{\hat{T}_{-1}^{-1}\hat{T}_{-1}^0\hat{T}_1^0\hat{T}_1^1}{(U+\Delta_0)^2(2U+\Delta_0)} - \frac{\hat{T}_{-1}^1\hat{T}_{-1}^0\hat{T}_1^0\hat{T}_1^{-1}}{(U-\Delta_0)^2(2U-\Delta_0)} - \frac{\hat{T}_{-1}^0\hat{T}_0^{-1}\hat{T}_0^1\hat{T}_1^0}{U^2(U+\Delta_0)} - \frac{\hat{T}_{-1}^0\hat{T}_0^1\hat{T}_0^{-1}\hat{T}_1^0}{U^2(U-\Delta_0)} -$$

$$-\frac{2\hat{T}_{-1}^{-1}\hat{T}_0^0\hat{T}_0^0\hat{T}_1^1 + \hat{T}_{-1}^{-1}\hat{T}_{-1}^{-1}\hat{T}_1^1\hat{T}_1^1 - 2\hat{T}_{-1}^{-1}\hat{T}_1^1\hat{T}_{-1}^{-1}\hat{T}_1^1}{2(U+\Delta_0)^3} - \frac{\hat{T}_{-1}^{-1}\hat{T}_0^0\hat{T}_0^1\hat{T}_1^0 + \hat{T}_{-1}^0\hat{T}_0^{-1}\hat{T}_0^0\hat{T}_1^1}{U(U+\Delta_0)^2} -$$

$$-\frac{2\hat{T}_{-1}^1\hat{T}_0^0\hat{T}_0^0\hat{T}_1^{-1} + \hat{T}_{-1}^1\hat{T}_{-1}^1\hat{T}_1^{-1}\hat{T}_1^{-1} - 2\hat{T}_{-1}^1\hat{T}_1^{-1}\hat{T}_{-1}^1\hat{T}_1^{-1}}{2(U-\Delta_0)^3} - \frac{\hat{T}_{-1}^1\hat{T}_0^0\hat{T}_0^{-1}\hat{T}_1^0 + \hat{T}_{-1}^0\hat{T}_0^1\hat{T}_0^0\hat{T}_1^{-1}}{U(U-\Delta_0)^2} -$$

$$-\frac{\hat{T}_{-1}^{-1}\hat{T}_{-1}^0\hat{T}_1^1\hat{T}_1^0 + \hat{T}_{-1}^0\hat{T}_{-1}^{-1}\hat{T}_1^0\hat{T}_1^1}{U(U+\Delta_0)(2U+\Delta_0)} - \frac{\hat{T}_{-1}^1\hat{T}_{-1}^0\hat{T}_1^{-1}\hat{T}_1^0 + \hat{T}_{-1}^0\hat{T}_{-1}^1\hat{T}_1^0\hat{T}_1^{-1}}{U(U-\Delta_0)(2U-\Delta_0)} - \frac{\hat{T}_{-1}^0\hat{T}_{-1}^{-1}\hat{T}_1^1\hat{T}_1^0}{U^2(2U+\Delta_0)} -$$

$$-\frac{\hat{T}_{-1}^0\hat{T}_{-1}^1\hat{T}_1^{-1}\hat{T}_1^0}{U^2(2U-\Delta_0)} - \frac{\hat{T}_{-1}^{-1}\hat{T}_1^1\hat{T}_{-1}^{-1}\hat{T}_1^1}{2U(U+\Delta_0)^2} - \frac{\hat{T}_{-1}^1\hat{T}_{-1}^1\hat{T}_1^{-1}\hat{T}_1^{-1}}{2U(U-\Delta_0)^2} + \frac{\hat{T}_{-1}^{-1}\hat{T}_{-1}^1\hat{T}_1^1\hat{T}_1^{-1} + \hat{T}_{-1}^1\hat{T}_{-1}^{-1}\hat{T}_1^{-1}\hat{T}_1^1}{2U(U^2-\Delta_0^2)} +$$

$$+\frac{2U+\Delta_0}{2U^2(U+\Delta_0)^2}\left[\hat{T}_{-1}^{-1}\hat{T}_1^1\hat{T}_{-1}^0\hat{T}_1^0 + \hat{T}_{-1}^0\hat{T}_1^0\hat{T}_{-1}^{-1}\hat{T}_1^1\right] + \frac{2U-\Delta_0}{2U^2(U-\Delta_0)^2}\left[\hat{T}_{-1}^1\hat{T}_1^{-1}\hat{T}_{-1}^0\hat{T}_1^0 +\right.$$

$$\left.+\hat{T}_{-1}^0\hat{T}_1^0\hat{T}_{-1}^1\hat{T}_1^{-1}\right] + \frac{U}{(U^2-\Delta_0^2)^2}\left[\hat{T}_{-1}^{-1}\hat{T}_1^1\hat{T}_1^1\hat{T}_1^{-1} + \hat{T}_{-1}^1\hat{T}_1^{-1}\hat{T}_{-1}^{-1}\hat{T}_1^1\right]. \quad (14)$$

**The Hubbard operators.** To proceed further it is convenient to introduce the so-called Hubbard operators $\hat{X}_i^{a \leftarrow b} \equiv |a\rangle_i{}_i\langle b|$ [18], which are defined on each site of the lattice and describe all possible transitions between the local basis states: unoccupied $|0\rangle_i$, singly occupied with "up"-spin $|\uparrow\rangle_i$, singly occupied with "down"-spin $|\downarrow\rangle_i$, and doubly occupied $|2\rangle_i$. The Hubbard operators which change the number of electrons on the site by an even or odd number are respectively Bose- or Fermi-like operators. They obey the following on-site multiplication rule

$$\hat{X}_i^{p \leftarrow q} \cdot \hat{X}_i^{r \leftarrow s} = \delta_{qr}\hat{X}_i^{p \leftarrow s} \quad (15a)$$

and the following commutation relations:

$$\left[\hat{X}_i^{p \leftarrow q}, \hat{X}_j^{r \leftarrow s}\right]_\pm = \delta_{ij}\left(\delta_{qr}\hat{X}_j^{p \leftarrow s} \pm \delta_{ps}\hat{X}_j^{r \leftarrow q}\right), \quad (15b)$$

where the upper sign stands for the case when both operators are Fermi-like, otherwise the lower sign should be adopted.

The original electron creation and annihilation operators can be expressed in terms of the Hubbard operators in the following way:

$$\hat{c}_{i\sigma}^+ = \hat{X}_i^{\sigma \leftarrow 0} + sign(\sigma)\hat{X}_i^{2 \leftarrow \bar{\sigma}}, \qquad \hat{c}_{i\sigma} = (\hat{c}_{i\sigma}^+)^+ = \hat{X}_i^{0 \leftarrow \sigma} + sign(\sigma)\hat{X}_i^{\bar{\sigma} \leftarrow 2}. \quad (16)$$

Using (15)-(16) and noting that the Hubbard operators at half-filling satisfy the relations

$$\hat{X}_i^{\sigma\leftarrow 0}|\Psi\rangle_L = 0, \quad \hat{X}_i^{\sigma\leftarrow 2}|\Psi\rangle_L = 0,$$

all the products of $\hat{T}_m^\alpha$ operators in the transformed Hamiltonian (14) can be rewritten in terms of the $\hat{X}_i^{\sigma\leftarrow\sigma'}$ Hubbard operators. Making use of Eqs. (16) once again, one can easily find that the spin $S = 1/2$ operators can also be expressed via the $\hat{X}_i^{\sigma\leftarrow\sigma'}$ operators in the following way

$$\hat{S}_i^z = \frac{1}{2}\sum_\sigma sign(\sigma)\hat{n}_{i\sigma} = \frac{1}{2}\left[\hat{X}_i^{\uparrow\leftarrow\uparrow} - \hat{X}_i^{\downarrow\leftarrow\downarrow}\right], \tag{17a}$$

$$\hat{S}_i^+ = \hat{c}_{i\uparrow}^+\hat{c}_{i\downarrow} = \hat{X}_i^{\uparrow\leftarrow\downarrow}, \quad \hat{S}_i^- = (\hat{S}_i^+)^+ = \hat{c}_{i\downarrow}^+\hat{c}_{i\uparrow} = \hat{X}_i^{\downarrow\leftarrow\uparrow}. \tag{17b}$$

Inverting (17), after some routine algebra, we finally obtain that, in the 4th order approximation, the strong-coupling effective spin Hamiltonian for the A$_2$B$_2$ Hubbard chain has the following form:

$$\hat{H}_{eff} = \sum_{i=1}^{N/2} J_i \hat{\vec{S}}_{2i-1} \cdot \hat{\vec{S}}_{2i} + J'\sum_{i=1}^{N} \hat{\vec{S}}_i \cdot \hat{\vec{S}}_{i+2}, \tag{18}$$

with

$$J_{2i} \equiv \frac{4t^2}{U}\left[1 - \frac{4t^2}{U^2}\right], \quad J_{2i+1} \equiv \frac{4t^2}{U}\cdot\frac{U^2}{U^2-\Delta_0^2}\left[1 - \frac{2t^2}{U^2}\cdot\frac{2U^2-\Delta_0^2}{U^2-\Delta_0^2}\right], \quad J' \equiv \frac{4t^4}{U^3}\cdot\frac{U^2}{U^2-\Delta_0^2}.$$

The Heisenberg chain with alternating spin exchange (18) is well studied and exhibits a gapped excitation spectrum for arbitrary (non-equal) values of the nearest-neighbour spin exchange constants [19]. Thus we conclude that, in marked contrast with the standard IHM, the Hubbard model with tetramerized (AABB-type) modulation of the ionic potential displays in the strong coupling regime the properties of an insulator with gapped spin excitation spectrum.

**Conclusion**. We have derived the effective spin Hamiltonian for the half-filled extended ionic-Hubbard model with tetramerized modulation of the ionic potential. We have shown that in the strong coupling regime, the low-energy properties of the model are described by the effective spin $S = 1/2$ Hamiltonian with alternating nearest-neighbour spin exchange.

**Acknowledgements**. It is our pleasure to thank D. Baeriswyl for fruitful discussions. This work has been supported by the GNSF grant No. ST06/4-018. M.M. also acknowledges the generous hospitality of the Department of Physics of the University of Fribourg, where part of the work has been done.